\UseRawInputEncoding
\documentclass[11pt, a4paper]{article}
\usepackage[ margin=1in]{geometry}
\usepackage[affil-it]{authblk}
\usepackage{setspace}
\usepackage[sort&compress,numbers]{natbib}
\usepackage[english]{babel}
\usepackage{graphicx}
\usepackage{mathptmx}
\usepackage{verbatim}
\usepackage{graphicx}
\usepackage{amsmath}
\usepackage{amssymb}
\usepackage{amsfonts}
\usepackage{mathrsfs}
\usepackage{amsmath} 
\usepackage{romannum}
\usepackage{color}
\usepackage{graphicx}
\usepackage{bm} 
\usepackage{amssymb}
\usepackage{xspace}
\usepackage{color}
\usepackage{float}
\usepackage{bbold}
\usepackage{tabu}
\usepackage{textcomp}
\usepackage{dcolumn}
\usepackage{graphicx}
\usepackage{dcolumn}
\usepackage{bm}
\usepackage{epstopdf}
\usepackage{xr}
\usepackage{amsmath}
\usepackage{color}
\usepackage{xcolor}
\usepackage[normalem]{ulem}
\usepackage{wasysym}
\usepackage{siunitx}
\usepackage{physics}
\usepackage{lineno}
\usepackage{gensymb}
\usepackage[T1]{fontenc}
\usepackage{booktabs}
\usepackage[section]{placeins}
\usepackage{textcomp}
\usepackage{caption}
\usepackage[colorlinks=true, linkcolor=black, urlcolor=blue, citecolor=blue]{hyperref}
\usepackage{blindtext}
\usepackage[absolute]{textpos}

\makeatletter
\newcommand{\Tns}{$T_\mathrm{n}^\mathrm{sys}$}
\newcommand{\flin}{$f_{\mathrm{res}}^{\mathrm{lin}}$}
\newcommand{\Gsys}{$G_\mathrm{sys}$}
\newcommand{\Tnsjpa}{$T_\mathrm{n}^{\mathrm{sys,Gr-JPA}}$}
\makeatother

\begin{document}
\captionsetup[figure]{labelfont={bf},labelsep=period,name={Fig.}}

\pagenumbering{arabic}
\title{\vspace{-2cm}\normalfont \textbf{Quantum noise limited microwave amplification using a graphene Josephson junction}}

\renewcommand\Authfont{\small}

\author[1]{Joydip Sarkar}
\author[1]{Kishor V. Salunkhe}
\author[1]{Supriya Mandal}
\author[1]{Subhamoy Ghatak}
\author[1]{\\Alisha H. Marchawala} 
\author[1]{Ipsita Das}
\affil[1]{Department of Condensed Matter Physics and Materials Science, Tata Institute of Fundamental Research, Homi Bhabha Road, Mumbai 400005, India.}

\author[2]{Kenji Watanabe}
\affil[2]{Research Center for Functional Materials,
National Institute for Materials Science, 1-1 Namiki, Tsukuba 305-0044, Japan.}

\author[3]{Takashi Taniguchi}
\affil[3]{International Center for Materials Nanoarchitectonics,
National Institute for Materials Science,  1-1 Namiki, Tsukuba 305-0044, Japan.}

\author[1*]{R. Vijay}
\author[1*]{\\Mandar M. Deshmukh}
\affil[*]{\textnormal{r.vijay@tifr.res.in, deshmukh@tifr.res.in}}

\date{}
\maketitle

\begin{abstract}
Josephson junctions (JJ) and their tunable properties, including their nonlinearities, form the core of superconducting circuit quantum electrodynamics (cQED)~\cite{devoret_superconducting_2013}.  In quantum circuits, low-noise amplification of feeble microwave signals is essential and the Josephson parametric amplifiers (JPA)~\cite{bergeal_phase-preserving_2010} are the widely used devices. The existing JPAs are based on Al-AlO$\bm{_\mathrm{x}}$-Al tunnel junctions realized in a superconducting quantum interference device geometry, where magnetic flux is the knob for tuning the frequency. Recent experimental realizations of $2$D van der Waals JJs~\cite{calado_ballistic_2015,lee_ultimately_2015} provide an opportunity to implement various cQED devices~\cite{wang_coherent_2019,schmidt_ballistic_2018,kroll_magnetic_2018} with the added advantage of tuning the junction properties and the operating point using a gate potential. While other components of a possible $2$D van der Waals cQED architecture have been demonstrated $\text{--}$ quantum noise limited amplifier, an essential component, has not been realized. Here we implement a quantum noise limited JPA, using a graphene JJ, that has linear resonance gate tunability of $3.5$ GHz. We report $24$ dB amplification with $10$ MHz bandwidth and $-130$ dBm saturation power; performance on par with the best single-junction JPAs~\cite{bergeal_phase-preserving_2010,hatridge_dispersive_2011}. Importantly, our gate tunable JPA works in the quantum-limited noise regime which makes it an attractive option for highly sensitive signal processing. Our work has implications for novel bolometers $\text{--}$  the low heat capacity of graphene together with JJ nonlinearity can result in an extremely sensitive microwave bolometer embedded inside a quantum noise-limited amplifier. In general, our work will open up exploration of scalable device architecture of $2$D van der Waals materials by integrating a sensor with the quantum amplifier.
\end{abstract}
\clearpage 

Josephson junctions are the key building block for the superconducting quantum devices, ranging from a wide variety of superconducting qubits~\cite{devoret_superconducting_2013} to parametric amplifiers~\cite{vijay_invited_2009,aumentado_superconducting_2020}. An oxide tunnel barrier-based JJs are commonly used as nonlinear inductors where the strength of the nonlinearity is tailored by tuning the oxide barrier properties, microwave current bias, or by magnetic flux. In recent years, semiconductor nanostructure-based JJs have provided an electrostatic control of the Josephson nonlinearity and have opened up new exciting possibilities such as superconducting qubit with proximitized $2$D electron gas ($2$DEG)~\cite{larsen_semiconductor-nanowire-based_2015,de_lange_realization_2015}, and  parity-protected qubit~\cite{larsen_parity-protected_2020} among others. 

\begin{figure*}
    \hspace*{0.6cm}
    \centering
    \includegraphics[width=15.5cm]{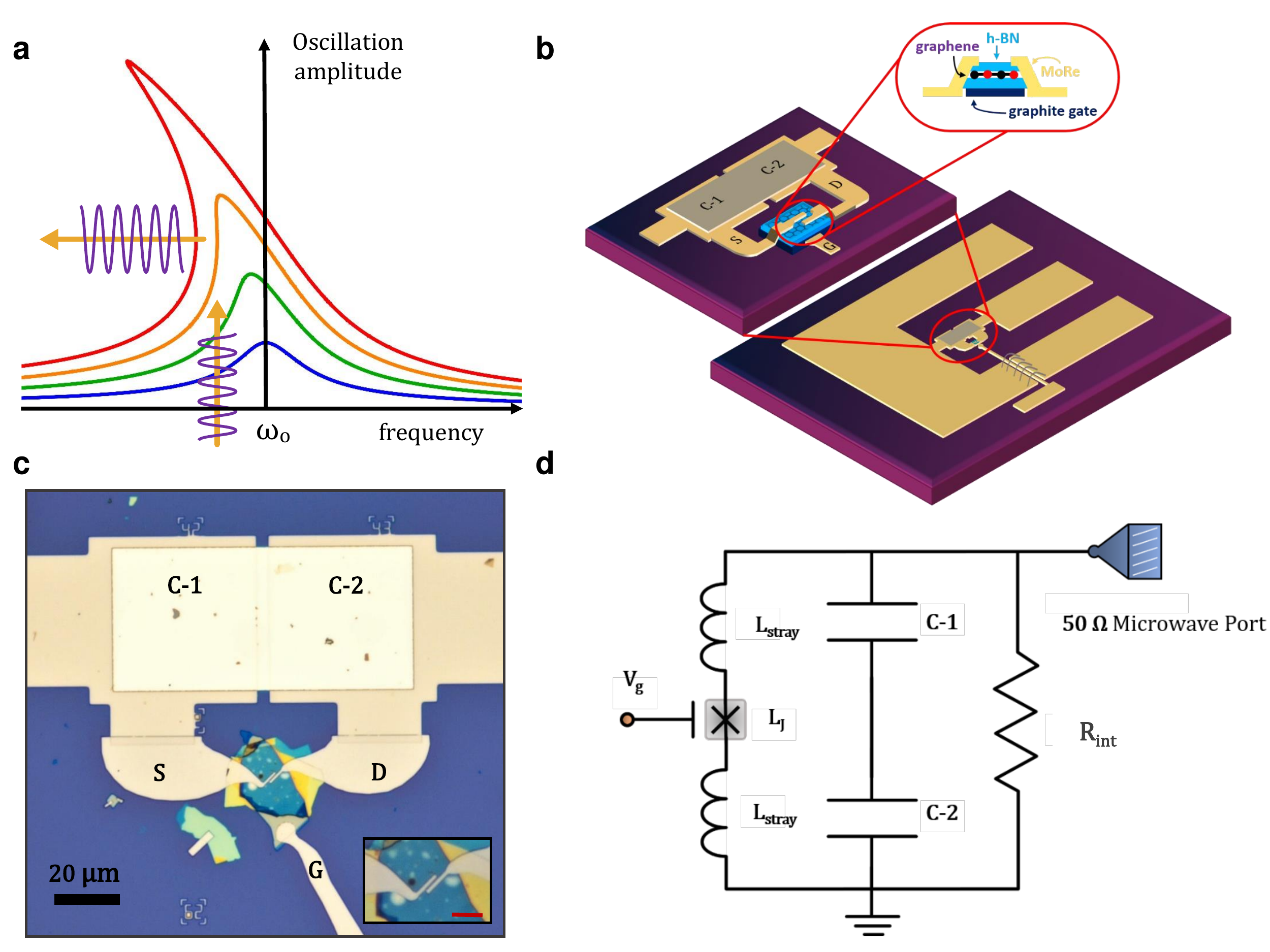}
    \caption{ \label{fig:fig1} {\textbf{Parametric amplification scheme and its implementation in a graphene Josephson junction (JJ) based LC resonator.} 
    \textbf{a,}~Schematic shows the amplitude response of a nonlinear Duffing oscillator as a function of frequency and increasing drive power (blue to red). With an increase in drive, the oscillation amplitude increases, and the nonlinear terms start to play a role. Consequently, the resonance peak starts to shift towards the left side. The sign of the first nonlinear term in the equation of motion of a Duffing oscillator determines the direction of this frequency shift. When the drive is increased at some point the amplitude response becomes very sharp (orange curve). This state of a Duffing oscillator is called the bifurcation regime where further increase in drive makes the oscillation amplitude a multi-valued function of frequency (red curve) and the oscillator becomes unstable. The orange curve is a bias point of the oscillator where it operates as a parametric amplifier that amplifies an input signal, as shown in the figure.
    \textbf{b,}~Schematic shows our experimental implementation of a Duffing oscillator. We make a coplanar waveguide (CPW) of MoRe, where the central line is terminated to the ground plane through a lumped element parallel LC resonator. We have two parallel plate MoRe-Al$_2$O$_3$-Al capacitors in series. We fabricate a MoRe-graphene-MoRe Josephson junction on an hBN-gr-hBN stack and patch the JJ to the capacitor leads, which results in a parallel LC resonator. We have a thick graphite flake below the hBN-gr-hBN stack which is connected to an electrode to implement a local back gating for carrier density modulation in gr-JJ.
    \textbf{c,}~Shows a zoomed optical micrograph of our device where the Josephson junction is in parallel to a series combination of two parallel plate capacitors C-$1$ and C-$2$. Both the capacitors have an identical dimension of $60\times60$ $\micro$m$^2$.
    In the image S, D, G implies respectively the source, drain, and gate electrodes of the gr-JJ in FET terminology. Inset shows a zoomed image of the junction where the scale bar is $5$ $\micro$m. The junction has length $\sim350$ nm and width $\sim4$ $\micro$m.
    \textbf{d,}~Shows the equivalent lumped element circuit model of the resonator where $L_{\mathrm{J}}$ is the junction inductance, $L_{\mathrm{stray}}$ is the stray inductance ($L_{\mathrm{stray}}=L_{\mathrm{geometric}}+L_{\mathrm{kinetic}}$) originating from the superconducting leads, C-$1$ and C-$2$ are the two parallel plate capacitors in series, $R_{\mathrm{int}}$ represents the internal loss of the resonator in the form of a resistor, and $V_{\mathrm{g}}$ is the applied gate potential to the gr-JJ. Eventually, the resonator is connected to a microwave port through a $50$ $\Omega$ matched environment for reflection-based measurements.}}

\end{figure*}

Probing the quantum states of a qubit requires measuring scattered photons in the cQED architecture. The typical single-photon limit experiments involve probe signals of ultra-low-power ($\sim-130$ dBm). Consequently, the output signals are too weak and comparable to the noise floor of any room temperature electronic detector. Hence, the output signals need sufficient amplification before processing. However, this increase in signal power comes at the price of added noise. Fundamental laws of quantum physics demand that any phase-preserving amplifier needs to add at least half a noise photon to the output signals in the high-gain limit. This bound on added noise is called the standard quantum limit (SQL). A linear amplifier amplifies an input signal as, $V_{\mathrm{out}}(t) =\sqrt{G}V_{\mathrm{in}}(t) + \epsilon(t)$, where $G$ is the real-valued amplifier power gain, $V_{\mathrm{in}}(t)$, $V_{\mathrm{out}}(t)$ are the input and output signals respectively, and $\epsilon(t)$ is the uncorrelated noise added by the amplifier. The noise spectral power is parameterized by the noise temperature ($T_\mathrm{n}$) of an amplifier. The Josephson parametric amplifiers (JPA)~\cite{aumentado_superconducting_2020,roy_broadband_2015} and Josephson parametric upconverters~\cite{schmidt_current_2020} are the two existing widely used devices that process signals near the standard quantum limit. There are different experimental realizations of parametric amplification in various cQED~\cite{ho_eom_wideband_2012,macklin_nearquantum-limited_2015}, optomechanical systems~\cite{massel_microwave_2011} in the form of a nonlinear Duffing oscillator. Fig.~\ref{fig:fig1}a shows the working scheme of a parametric amplifier, which under correct biasing condition amplifies an input signal. In cQED architecture, the JPAs utilize the intrinsic Kerr nonlinearity of a Josephson junction as the source of nonlinearity~\cite{vijay_invited_2009}. 

Recent experimental realizations of superconductor-normal-superconductor (SNS) junctions using various $2$D van der Waals materials provide a platform to devise gate tunable JJs. One can tune the junction properties like critical current ($I_{\mathrm{c}}$) hence junction inductance ($L_{\mathrm{J}}$) by electrostatic gating in a field-effect transistor-based device geometry. The gate tuning also provides critical control over the current phase relationship (CPR)~\cite{haller_phase-dependent_2022} that determines the extent of non-linearity. Recent experiments have used graphene JJs (gr-JJ) in various cQED devices including superconducting qubits~\cite{wang_coherent_2019}, microwave resonators~\cite{schmidt_ballistic_2018,kroll_magnetic_2018,dou_microwave_2021}. Because of the nonlinear inductive nature of JJs, the SNS junctions are potential candidates to be used in gate tunable JPAs. Having an extra control knob of gating makes these $2$D materials-based JPAs ideal for applications such as quantum amplitude sensing~\cite{stehlik_fast_2015,joas_quantum_2017,eddins_high-efficiency_2019}, single-shot qubit readout~\cite{mallet_single-shot_2009}, dispersive magnetometry~\cite{hatridge_dispersive_2011} and detection of dark matter axions~\cite{braine_extended_2020}. Integration of $2$D materials with unique electrical and thermal properties have further provided an avenue to explore novel detectors like the single-photon bolometers~\cite{walsh_graphene-based_2017,lee_graphene-based_2020,kokkoniemi_bolometer_2020,walsh_josephson_2021}. Here we report for the first time, the observation of quantum-limited microwave amplification in a graphene-based Josephson parametric amplifier (Gr-JPA), where we implement a gr-JJ to form the nonlinear LC resonator.

We fabricate the JJ on a few-layer graphene flake encapsulated between two flakes of hexagonal boron nitride (h-BN) and edge contacted with a type-$\mathrm{II}$ superconductor molybdenum-rhenium (MoRe) to realize the MoRe-graphene-MoRe planar SNS junction~\cite{wang_one-dimensional_2013,calado_ballistic_2015}. We start with a pre-patterned coplanar waveguide (CPW) made of MoRe, where we make two MoRe-Al$_2$O$_3$-Al parallel plate capacitors of identical dimensions ($60\times60$ $\micro$m$^2$) in series. At the last fabrication step, a JJ is fabricated on the hBN-graphene-hBN (hBN-gr-hBN) stack. The JJ is patched parallel to the capacitors to realize a parallel LC oscillator. We use a thick graphite flake below the hBN-gr-hBN stack to set the local back gate potential of the gr-JJ (see methods for details of the fabrication). Fig.~\ref{fig:fig1}b shows a schematic representation of the whole device on a SiO$_\mathrm{2}$/Si substrate. Fig.~\ref{fig:fig1}c shows an optical micrograph of the device, the junction has length $\sim350$ nm and width $\sim4$ $\micro$m. For microwave frequencies, the effective lumped element circuit model is shown in Fig.~\ref{fig:fig1}d. In series with $L_{\mathrm{J}}$ there is a finite stray inductance ($L_{\mathrm{stray}}$) present in the circuit, which is a combined contribution of the geometric and kinetic inductance of the superconducting leads. Minimizing $L_{\mathrm{stray}}$ in comparison to $L_{\mathrm{J}}$ is crucial for the stable operation of the JPA as we will discuss later in this manuscript.

\begin{figure*}

    \centering
	\includegraphics[width=15.5cm]{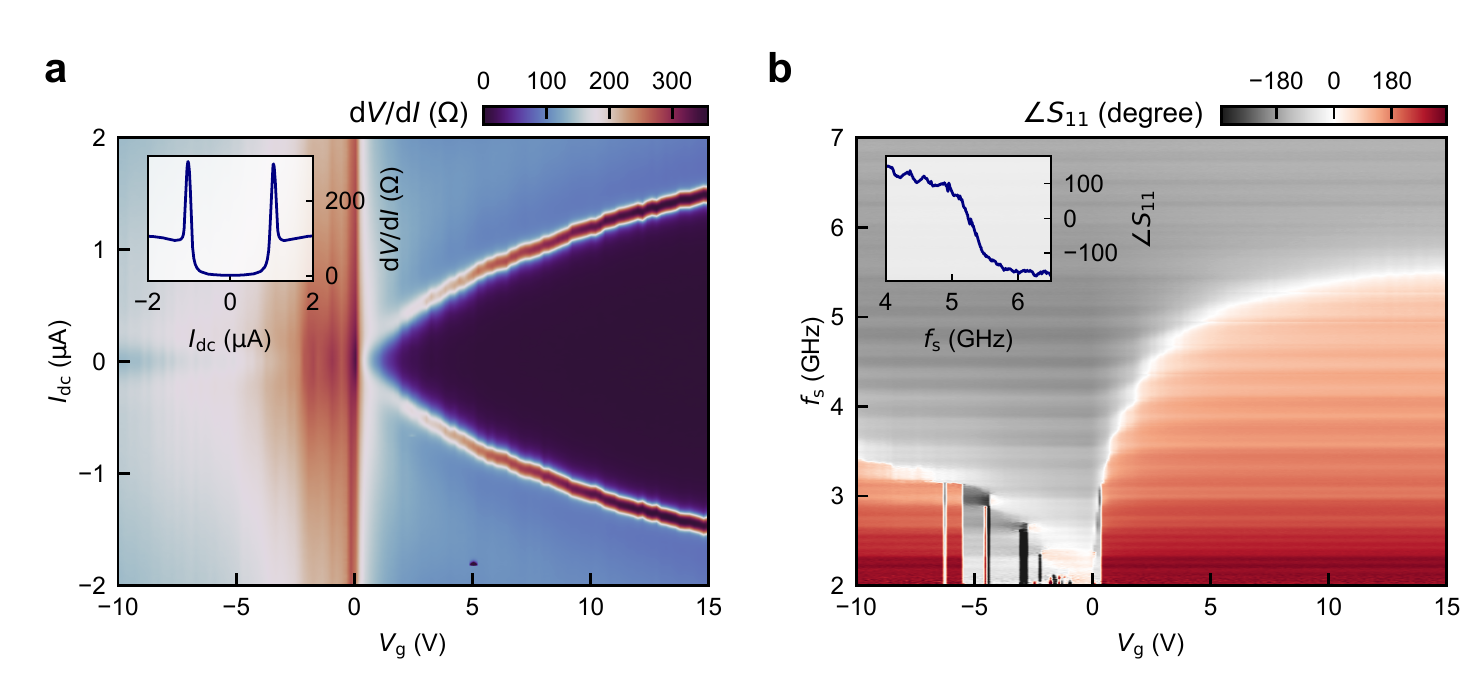}
    \caption{ \label{fig:fig2}{\textbf{Gate tunability of the critical current ($I_{\mathrm{c}}$) and junction inductance ($L_{\mathrm{J}}$).} \textbf{a,}~Shows a $4$-probe differential resistance ($\mathrm{d}V/\mathrm{d}I$) map of the gr-JJ as function of dc-bias current ($I_{\mathrm{dc}}$) and applied gate voltage ($V_{\mathrm{g}}$), measured at $20$ mK temperature. The dark blue region in the resistance map indicates the superconducting state of the junction where it has zero resistance. We extract the critical current ($I_{\mathrm{c}}$) value from the boundary of the sharp peak in the normal state resistance. The critical current becomes zero near the charge neutrality point (CNP) of graphene ($V_{\mathrm{g}}\sim0$ V) and increases up to $1.5$ $\micro$A as the doping is increased on the electron side. There is a strong asymmetry of $I_{\mathrm{c}}$ with respect to gating. Primarily we attribute this to the low transparency factor for holes arising from device fabrication. The inset shows a line slice of the differential resistance map at $V_{\mathrm{g}}=8$ V.
    \textbf{b,}~Shows reflected phase ($\angle S_{11}$) of the resonator plotted as a function of signal frequency ($f_{\mathrm{s}}$) and applied gate voltage ($V_{\mathrm{g}}$). Since the junction inductance ($L_{\mathrm{J}}$) is related to its critical current ($I_{\mathrm{c}}$) as Eq.~\eqref{eqn:m2}, the linear resonance frequency of the resonator (\flin) gets modulated as a function of gating. $V_{\mathrm{g}}$ tunes the linear resonance in a large frequency band of $2$ GHz to $5.5$ GHz. The $2\pi$ phase change at resonance indicates that the resonator is over coupled $(Q_{\mathrm{int}}\gg Q_{\mathrm{ext}})$ all along the electron doping side, which means the internal loss is less than external loss. 
    There is a frequency band of critically/under coupled region between $V_{\mathrm{g}}$=$-6$ to $0$ V where the resonator is lossy. This can be understood from the DC resistance map (Fig.~\ref{fig:fig2}a) as the junction has very weak superconducting proximity (low $I_{\mathrm{c}}$) in that gate voltage region. The inset shows a line slice of the phase plot at $V_{\mathrm{g}}=8$ V.
}}
\end{figure*}

Prior to microwave measurements, we perform a few basic DC characterizations of the junction to know the critical current ($I_{\mathrm{c}}$) value and its tunability with electrostatic doping. Fig.~\ref{fig:fig2}a shows a $4$-probe differential resistance ($\mathrm{d}V/\mathrm{d}I$) map of the junction as a function of dc-bias current ($I_{\mathrm{dc}}$) and applied gate voltage ($V_{\mathrm{g}}$), measured at $20$ mK temperature in a dilution fridge using standard lock-in detection technique (see methods for details of the setup). The dark blue region in the resistance map indicates the superconducting state of the junction where it has zero resistance and the boundary of the sharp peak in the normal state resistance indicates the value of the critical current ($I_{\mathrm{c}}$). The critical current is zero near the charge neutrality point (CNP) of graphene ($V_{\mathrm{g}}\sim0$ V) and increases up to $1.5$ $\micro$A as the electrostatic doping is increased on the positive $V_{\mathrm{g}}$ side. There is a strong asymmetry of $I_{\mathrm{c}}$ on the electron and hole sides which is similar to the previously reported results on gr-JJs~\cite{calado_ballistic_2015,schmidt_ballistic_2018}. We attribute this primarily to the low transparency factor for holes arising from device fabrication-related steps. We also measure the Fraunhofer modulation of $I_{\mathrm{c}}$ as a function of the applied magnetic field (see supplementary Fig. S7). Following past work~\cite{titov_josephson_2006} on modeling the supercurrent in gr-JJs, we model the CPR using a simplified relation,
\begin{equation}\label{eqn:m1}
    I_{\mathrm{s}}(\phi) = \frac{\pi\Delta_0}{2eR_{\mathrm{n}}} \frac{\sin{\phi}}{\sqrt{1-\tau\sin^2{(\phi/2)}}}
\end{equation}
based on the formation of Andreev bound states in the normal metal region. In Eq.~\eqref{eqn:m1}, $I_{\mathrm{s}}$ is the supercurrent, $\phi$ is the phase difference across the junction, $\Delta_0$ is the induced superconducting energy gap in graphene, $R_{\mathrm{n}}$ is the normal state resistance of the junction and $\tau$ is the averaged transparency factor for $N$ conducting channels in graphene. We use the CPR shown in Eq.~\eqref{eqn:m1} to numerically extract the evolution of transparency ($\tau$) as a function of $V_{\mathrm{g}}$ (see supplementary Fig. S8). Given a CPR one can estimate the junction inductance ($L_{\mathrm{J}}$) as,
\begin{equation}\label{eqn:m2}
    L_{\mathrm{J}}(\phi) = \frac{\hslash}{2e} \left(\frac{\partial I_{\mathrm{s}}}{\partial\phi}\right)^{-1}.
\end{equation}
Combining, Eq.~\eqref{eqn:m1} and Eq.~\eqref{eqn:m2} we numerically calculate the nonlinear junction inductance of the form $L_{\mathrm{J}}(I)=L_0+L_1I+L_2I^2+L_3I^3+..$ where $L_i^{~,}$s are the nonlinear inductor coefficients and $I$ is the drive current. We use this form of $L_{\mathrm{J}}(I)$ to simulate a nonlinear phase response of the resonator which we discuss later in this manuscript. In the next section, we discuss the microwave response of our Gr-JPA.

The value of the $I_{\mathrm{c}}$ based on DC measurements gives us an estimation of the junction inductance. The device parameters were chosen to target the maximum resonant frequency around $6$ GHz. We probe the resonator with low-power microwave signals and measure the reflected signals using a vector network analyzer (VNA). The measurements were done in a dilution fridge at $40$ mK temperature (see supplementary section-I for details on the setup). In Fig.~\ref{fig:fig2}b, the reflected phase ($\angle S_{11}$) of the resonator is plotted as a function of signal frequency ($f_{\mathrm{s}}$) and applied gate voltage ($V_{\mathrm{g}}$). As the junction inductance ($L_{\mathrm{J}}$) is a function of electrostatic doping, the linear resonance frequency (\flin) gets modulated from $2$ GHz up to $5.5$ GHz with gating; this wide frequency tuning is a key aspect of our device. The $2\pi$ phase change at resonance is the indication of over-coupled $(Q_{\mathrm{int}}\gg Q_{\mathrm{ext}})$ state of the resonator, which means the internal loss rate of the resonator is less than the external loss rate. The resonator is over-coupled all along on the positive $V_{\mathrm{g}}$ side. However, there is a frequency band of critically/under-coupled region between $V_{\mathrm{g}}$=$-6$ to $0$ V where the resonator is lossy. This can be understood from the DC resistance map (Fig.~\ref{fig:fig2}a) as the junction has very weak superconducting proximity (low $I_{\mathrm{c}}$) in that gate voltage region. These lumped element resonators are by construction low $Q$ in nature, which is desired to get a large gain-bandwidth product as discussed later in this manuscript. 

\begin{figure*}
    \centering
	\includegraphics[width=16cm]{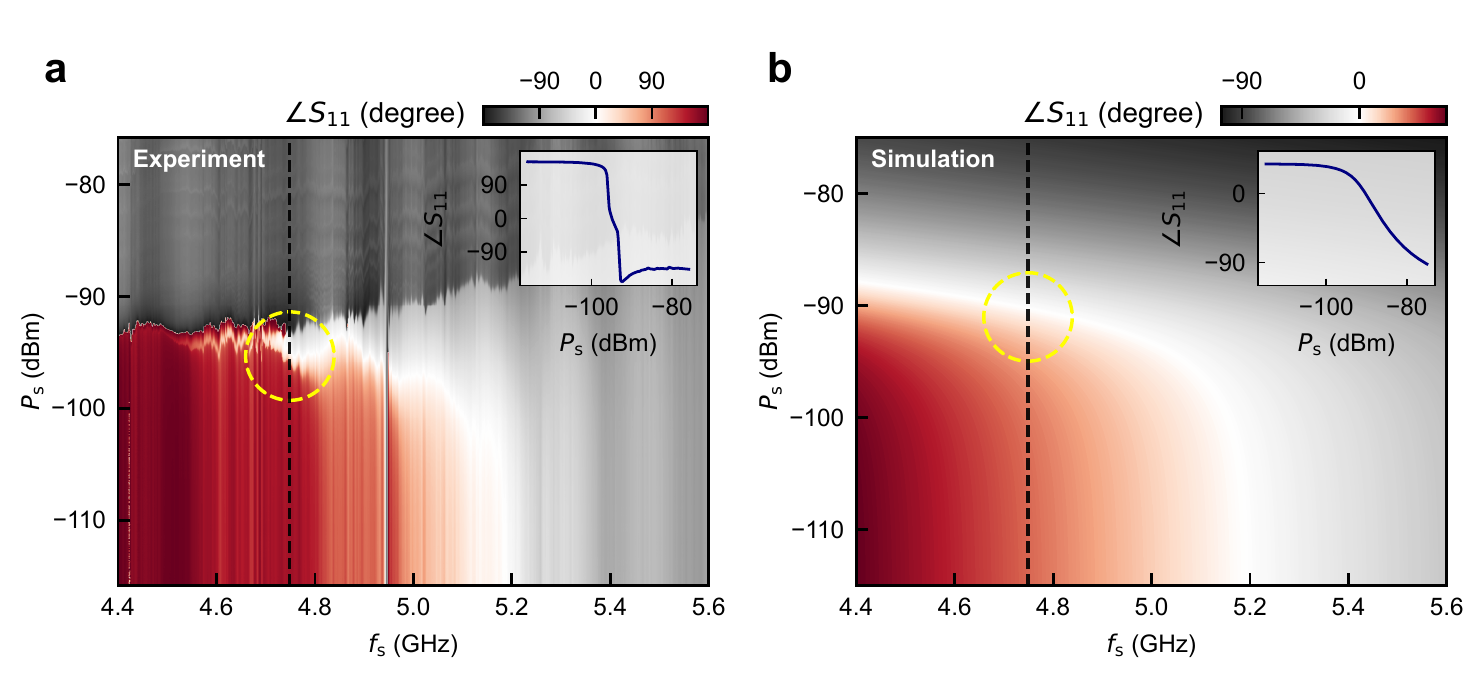}
    \caption{ \label{fig:fig3} {\textbf{Nonlinear phase diagram (experimental and simulated).} 
    \textbf{a,}~Shows reflected phase ($\angle S_{11}$) of the resonator plotted as a function of microwave signal frequency ($f_{\mathrm{s}}$) and power ($P_{\mathrm{s}}$). Here we fix the $V_{\mathrm{g}}$ at $4$ V, which fixes the linear resonance (\flin) at $\sim5.185$ GHz at low enough power ($-115$ dBm). However, with increasing signal power the resonator moves into the nonlinear regime and the resonance starts to shift towards the left side of linear resonance. Resonance shifting towards the left is the indication of the negative sign of the first nonlinear term in the SNS junction CPR. An increase in power pushes the resonator towards the bifurcation regime beyond which further increase in power makes the resonator unstable. We mark a frequency point at $4.749$ GHz near the bifurcation regime (yellow dashed circle) where we observe parametric amplification while introducing a pump tone at that frequency; this result is discussed in detail in Fig.~\ref{fig:fig4}. The inset shows a line slice of the phase response as a function of signal power at $4.749$ GHz. The amplification is expected to occur at the power where the phase falls sharply as indicated in the inset.
    \textbf{b,}~Shows a simulated nonlinear phase diagram at the same bias point of $V_{\mathrm{g}}=4$ V. We see resonance shifting with increasing signal power which qualitatively captures the main feature of a nonlinear resonator and agrees well with the experimental data. The simulation is performed in Microwave Office based on a nonlinear inductor model ($L_{\mathrm{J}}(I)=L_0+L_1I+L_2I^2+L_3I^3+..$) that we construct using Eq.~\eqref{eqn:m1} and Eq.~\eqref{eqn:m2}.
    }}
\end{figure*}

To explore the nonlinear aspects of the resonator, we fix the gate voltage $V_{\mathrm{g}}$ at $4$ V which sets the linear resonant frequency at $5.185$ GHz. Fig.~\ref{fig:fig3}a shows the experimental nonlinear phase diagram, where we measure the reflected phase ($\angle S_{11}$) of the resonator as a function of microwave signal frequency ($f_{\mathrm{s}}$) and power ($P_{\mathrm{s}}$). At low powers, we observe the linear behavior with a smooth $2\pi$ phase shift where a phase of zero (white) marks the effective resonant frequency. As the drive power is increased, we observe the resonant frequency shift to lower values as expected for an oscillator with a CPR shown in Eq.~\eqref{eqn:m1}. We also observe that the phase shift through resonance becomes sharper which indicates the onset of nonlinear behavior. The inset of Fig.~\ref{fig:fig3}a shows a line slice of the phase response as a function of signal power at a frequency of $4.749$ GHz which also shows a sharp transition. This is another characteristic feature of a driven nonlinear resonator where one can cross the effective resonant frequency by changing either drive frequency or power. At higher drive powers, the oscillator first becomes bistable and then unstable~\cite{vijay_invited_2009}. However, the region of interest for observing parametric amplification is shown in the inset (Fig.~\ref{fig:fig3}a), where the response is stable and shows a phase change with drive power. Fig.~\ref{fig:fig3}b shows a numerically simulated phase diagram at the same bias point of $V_{\mathrm{g}}=4$ V, which qualitatively captures the main features of a driven nonlinear resonator and agrees well with the experimental data (simulations details are provided in supplementary section-V).

\begin{figure*}
    \centering
	\includegraphics[width=15cm]{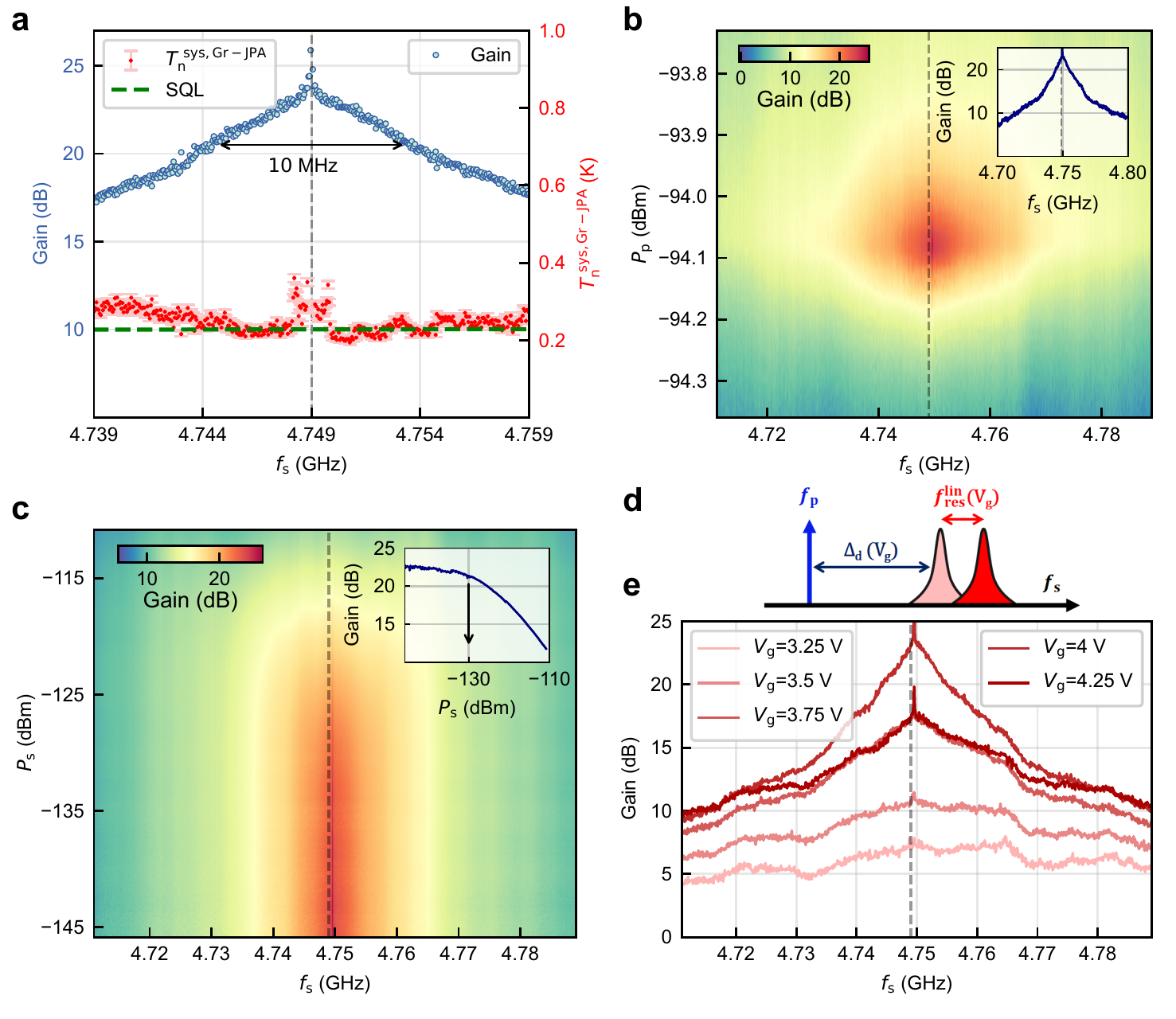}
    \caption{ \label{fig:fig4} {\textbf{Gr-JPA as a quantum limited amplifier.} 
    \textbf{a,}~Shows parametric amplification (left axis) at a bias point of $V_{\mathrm{g}}=4$ V. $V_{\mathrm{g}}$ fixes the linear resonance (\flin) at $\sim5.185$ GHz. Based on the nonlinear phase diagram (Fig.~\ref{fig:fig3}a) we add a pump tone near the bifurcation regime. Carefully adjusting the pump frequency ($f_{\mathrm{p}}$) and pump power ($P_{\mathrm{p}}$) we see $24$ dB amplification of signals around the pump frequency having $\sim10$ MHz amplification bandwidth. Here we set $f_{\mathrm{p}}$ at $4.749$ GHz and $P_{\mathrm{p}}$ at $-94.08$ dBm. The measured system noise temperature of the amplification chain (\Tnsjpa) including the Gr-JPA, shown on the right axis is very close to the standard quantum limit (SQL). The frequency dependent SQL is plotted in the green dashed line ($T_\mathrm{n}^{\mathrm{SQL}}=hf_{\mathrm{s}}/k_{\mathrm{B}}$). The \Tnsjpa being close to SQL indicates the quantum-limited noise performance.
    \textbf{b,}~Shows the gain evolution as a function of probe signal frequency ($f_{\mathrm{s}}$) and pump power ($P_{\mathrm{p}}$). The maximum gain region roughly follows an ellipse-like domain beyond which the gain starts to drop rapidly. The inset shows a line slice of the same gain plot as a function of signal frequency ($f_{\mathrm{s}}$) at a fixed pump power $-94.08$ dBm.
    \textbf{c,}~Shows the $1$-dB compression point characterization of the amplifier. Here we plot gain as a function of probe signal frequency ($f_{\mathrm{s}}$) and power ($P_{\mathrm{s}}$). The maximum gain is stable and constant for low powers; with the increase in signal power, the maximum gain starts to drop. 
    The inset shows the $1$-dB compression point of $\sim-130$ dBm at a signal frequency $4.748$ GHz.
    \textbf{d,~e,}~Shows frequency detuning effect on gain. Here we fix the pump tone ($f_{\mathrm{p}}$) and change the detuning between the pump and linear resonance frequency ($\Delta_{\mathrm{d}}=f_{\mathrm{res}}^{\mathrm{lin}}-f_{\mathrm{p}}$) by pushing the linear resonance up in frequency by increasing $V_{\mathrm{g}}$. As the $\Delta_{\mathrm{d}}$ increases the gain starts to increase and eventually reaches a maximum. However, further increase in detuning drives the resonator into the bistable region and the gain starts to drop.
    }}
\end{figure*}

We now discuss the central results of this manuscript demonstrating parametric amplification. At the same gate bias ($V_{\mathrm{g}}=4$ V) discussed above, we add a pump tone at $f_{\mathrm{p}} = 4.749$ GHz using a separate microwave source and set the pump power to bias the resonator at the sharp transition shown in the inset of Fig.~\ref{fig:fig3}a. A second weaker (by $\sim 40-50 $ dB) probe tone is added using the VNA to measure the amplification. We extract the gain by comparing the output at the probe frequency with and without the pump tone turned on. We then optimize the gain by a fine adjustment of the pump frequency and pump power. At the optimal pump frequency and power, we observe parametric amplification with $24$ dB gain in a $\sim10$ MHz bandwidth as shown in Fig.~\ref{fig:fig4}a. The noise properties of the Gr-JPA are calculated by measuring the signal-to-noise improvement in the presence of the Gr-JPA and using the system noise temperature of the rest of the amplification chain measured in a separate experiment (see supplementary section-II). The total system noise temperature (\Tnsjpa) including the Gr-JPA is plotted in Fig.~\ref{fig:fig4}a (right axis) and shows quantum-limited noise performance. The evolution of the gain curve with pump power ($P_{\mathrm{p}}$) is shown in Fig.~\ref{fig:fig4}b and we observe that the gain is maximized at a particular value of pump power. This point is the preferred bias point for amplifier operation as the gain remains stable against small drifts in the pump power.
The maximum gain achievable at a particular pump frequency depends on the detuning between the pump and linear resonance frequency~\cite{vijay_invited_2009} with larger detunings providing larger gain as shown in Fig.~\ref{fig:fig4}e. Here we fix the pump tone ($f_{\mathrm{p}}$) and change the detuning ($\Delta_{\mathrm{d}}=f_{\mathrm{res}}^{\mathrm{lin}}-f_{\mathrm{p}}$) between the pump and linear resonance frequency by pushing the linear resonance up in frequency by increasing $V_{\mathrm{g}}$, shown in the schematic Fig.~\ref{fig:fig4}d. As $\Delta_{\mathrm{d}}$ increases the gain starts to increase, reaches a maximum as it approaches the threshold for critical behavior, and then drops rapidly. The variation of the gain with pump power, pump detuning, and signal frequency described above confirms that our Gr-JPA shows the nonlinear behavior expected from its CPR.

The observed $24$ dB gain along with quantum-limited noise temperature demonstrates that our device can be used as an amplifier in quantum circuits. Another important characteristic of an amplifier is the $1$-dB compression point which we show in  Fig.~\ref{fig:fig4}c. We plot the amplifier gain curve for increasing signal power ($P_{\mathrm{s}}$) to observe the power at which the gain drops by $1$-dB from its maximum value and obtain a value of $-130$ dBm. This is on par with the best single oscillator based JPAs using SIS junctions \cite{bergeal_phase-preserving_2010,hatridge_dispersive_2011,aumentado_superconducting_2020}. We also measure the amplifier performance at different $V_{\mathrm{g}}$ and obtain at least $20$ dB maximum gain (see supplementary Fig. S5).

In addition, for the stable operation of the JPA a careful optimization of the device parameters, stray inductance ($L_{\mathrm{stray}}$) and capacitance ($C$), is crucial. The gain-bandwidth product of a conventional JPA follows the relation $B\sqrt{G}\propto1/Q$. Having a low $Q$ resonator helps in increasing the gain-bandwidth product. However, one cannot reduce $Q$ arbitrarily in order to increase both the gain and bandwidth together. With a low $Q$ the JPA needs to operate at high pump powers; hence with high power, higher-order nonlinear processes become important and prevent the stable operation of the JPA. An approximate condition for the stability is $Qp$ $\gtrsim$ $5$~\cite{Macklin_PhD_thesis_2007}, where $p$ = $L_{\mathrm{J}}$/$L_{\mathrm{tot}}$ is the participation ratio of the Josephson inductance to the total inductance. There is always some stray inductance present in the circuit as discussed earlier which constraints $p<1$. In our Gr-JPA, at the operating point, we estimate $Q\approx11$ and $p\approx0.93$ which makes the product $Qp\approx10$.

In summary, we report $24$ dB quantum-limited amplification in a gate tunable graphene JPA. Our work is the first implementation of a $2$D van der Waals-based parametric amplifier and is an essential building block for the implementation of quantum processors based on van der Waals architecture. Recent advances have demonstrated the realization of microwave circuits~\cite{schmidt_ballistic_2018,kroll_magnetic_2018}, transmon qubit~\cite{wang_coherent_2019} and compact low-loss quantum devices~\cite{wang_hexagonal_2022} by leveraging the properties of van der Waals materials; our work augments efforts to use the hybrid architecture of $2$D materials for quantum processors by demonstrating an amplifier. Further, harnessing the extremely sensitive nature of JPA nonlinearity, and low specific heat of graphene will enable the design of integrated photon detectors. The spatially extended nature of the gr-JJs, in contrast with Al-AlO$_\mathrm{x}$-Al tunnel junctions, allow facile coupling between magnetic materials and the Gr-JPA for realizing a dispersive magnetic sensor.

\section*{Methods:}
\renewcommand{\thesection}{\Roman{section}}
\setcounter{section}{0}
\section{Device fabrication}
We make multiple Gr-JPA devices. Our fabrication process involves four major steps. First, we fabricate a CPW of MoRe on a SiO$_2$/Si substrate following standard e-beam lithography. The CPW is designed to have a characteristic impedance ($Z_0$) of $50~\Omega$. The MoRe films are deposited using DC magnetron sputtering in a high vacuum chamber ($\sim4\times10^{-7}$ mbar); with sputtering pressure $\sim2\times10^{-3}$ mbar. Second, we make two MoRe-Al$_2$O$_3$-Al parallel plate capacitors (main Fig.1c). The Al$_2$O$_3$ and Al films are deposited using angled uniform rotational e-beam evaporation to ensure uniform film deposition in a high vacuum environment ($\sim2\times10^{-7}$ mbar). The Al$_2$O$_3$ dielectric thickness is $\sim50$ nm. Third, while the previously discussed steps are in process we prepare a few hBN-gr-hBN stacks. We exfoliate flakes using scotch tape mechanical exfoliation of bulk crystals of graphene and hBN; next, we choose a few-layer graphene flake (mostly bilayer/trilayer), $20-30$ nm thick top, and $50-60$ nm thick bottom hBN flakes. We then stack the flakes one by one using PC (Poly(bisphenol A carbonate))/ PDMS (polydimethylsiloxane) stamps. Once we have the CPW with capacitor and hBN-gr-hBN stack ready, we exfoliate graphite on PDMS and transfer a thick graphite flake on the CPW with capacitor chip for back gating. Next, we drop the hBN-gr-hBN stack on the gate graphite flake and coat the chip with PMMA $495$ A$4$ + PMMA $950$ A4 bilayer e-beam resists with a combined resist thickness of $\sim350$ nm. Fourth, we do standard e-beam lithography followed by CHF$_3$/O$_2$ reactive ion etching to define the edge contacts for the JJ. Prior to MoRe deposition, we do an in-situ argon (Ar) plasma cleaning inside the sputtering chamber. This in-situ Ar plasma cleaning improves the contact transparency of the JJs significantly. Finally, we deposit $\sim60$ nm MoRe to get the contacts. Once the fabrication is done we place the device chip on a circuit board and wire-bond for measurements.

\section{Measurement technique}
We perform low-temperature DC and microwave measurements on our Gr-JPA to characterize its DC and microwave properties. The DC and microwave measurements were done in two different dilution fridges at base temperatures of $20$ mK and $40$ mK respectively; the measurement techniques are discussed below
\subsection{DC measurements:}
The DC measurements primarily involve the $4$-probe differential resistance measurements of the JJs using standard lock-in detection technique with SRS 830 lock-in amplifier. We send small ($\sim50$ nA$\ll I_\mathrm{c}$) low frequency ($17$ Hz) current through the parallel LC resonator and measure the voltage across it as a function of DC bias current ($I_\mathrm{dc}$) and applied back gate voltage ($V_\mathrm{g}$); due to the frequency of the AC current being very small the capacitor can effectively be thought as open. Hence the voltage across the parallel LC we assume to be the response of the JJ only. We take precautions to filter any undesired high-frequency signals which may travel from room temperature to the device. Three-stage filtering of the DC lines is done using low pass RC filters that are kept at different temperature plates of our dilution fridge (room temperature, $4$ K plate, $20$ mK plate) and they all have cut-off frequencies $\sim70$ kHz. The line used for the back gate ($V_\mathrm{g}$) also passes through the aforementioned three-stage RC filtering; additionally, we put a $10$ Hz low pass RC filter at room temperature. In addition to the RC filters we also have copper powder filters (at $20$ mK) and eccosorb filters (at $4$ K) in our measurement lines to attenuate any higher frequency noise.
\subsection{Microwave measurements:}
The microwave measurements involve the standard reflectometry technique. Similar to DC setup in microwave measurements we take precautions for filtering any undesired noise that may reach the device. The microwave set up along with the wiring diagram is discussed in detail in supplementary section-I.
\section{Noise calibration and measurements}
To extract the noise temperature of the Gr-JPA, we first measure the frequency-dependent system noise temperature (\Tns) of our amplification chain using a shot noise tunnel junction (SNTJ) as a calibrated source of the noise. The calibration of SNTJ and Gr-JPA noise measurements are discussed in detail in the supplementary section-II.


\section*{Acknowledgements:}
We thank Vibhor Singh, Sophie Gueron, Ziwei Dou, Helene Bouchiat, Pratap C. Adak, Subhajit Sinha, Sanat Ghosh, Sumeru Hazra for helpful discussions and comments. We thank Jhuma Saha, SLD Varma, Krishnendu Maji for experimental assistance. We acknowledge Nanomission grant SR/NM/NS-45/2016 and DST SUPRA SPR/2019/001247 grant along with Department of Atomic Energy of Government of India 12-R\&D-TFR-5.10-0100 for support.
Preparation of hBN single crystals is supported by the Elemental Strategy Initiative conducted by the MEXT, Japan (Grant Number JPMXP0112101001) and  JSPS KAKENHI (Grant Numbers 19H05790 and JP20H00354).

\section*{Author Contributions:}
J.S. fabricated the devices. J.S. and K.V.S. did the measurements and analyzed the data. S.G., A.H.M., I.D., and S.M. assisted in developing the device fabrication method and experimental setup. K.W. and T.T. grew the hBN crystals. R.V. led the microwave measurements. J.S., K.V.S., R.V., and M.M.D. wrote the manuscript with inputs from everyone. M.M.D. supervised the project.

\clearpage

\clearpage

\begin{center}
\textbf{\huge Supplementary Information}
\end{center}
\renewcommand{\thesection}{\Roman{section}}
\setcounter{section}{0}
\renewcommand{\thefigure}{S\arabic{figure}}
\captionsetup[figure]{labelfont={bf},name={Supplementary Fig.}}
\setcounter{figure}{0}
\renewcommand{\theHequation}{Sequation.\theequation}
\renewcommand{\theequation}{S\arabic{equation}}
\setcounter{equation}{0}

\section{Microwave measurement setup}
\begin{figure}[H]
    \hspace*{0.3cm}
    \centering
	\includegraphics[width=0.85\linewidth]{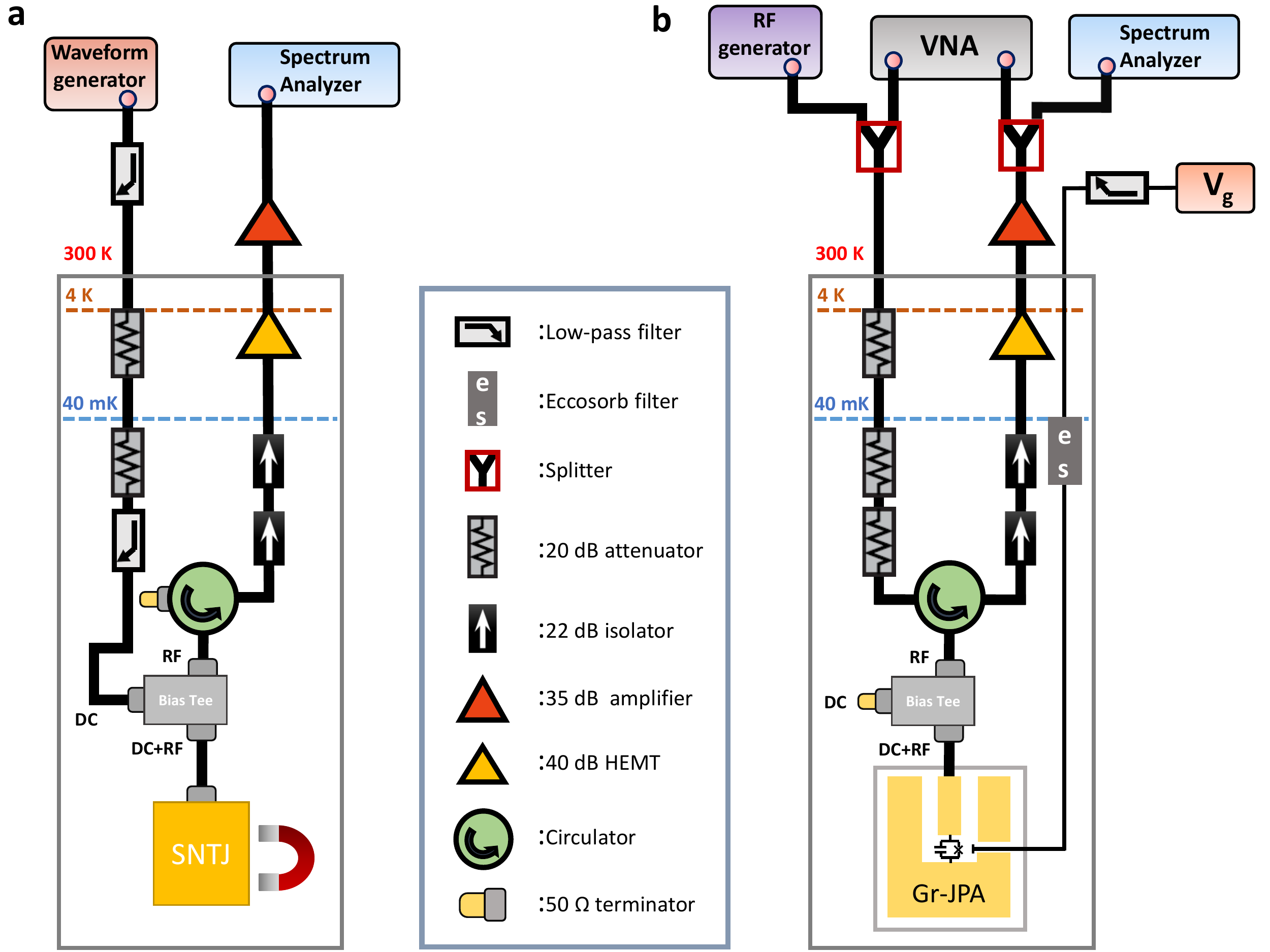}
    \caption{\label{fig_1} {\textbf{Microwave setup for SNTJ calibration and Gr-JPA noise measurements.}
    \textbf{a,}~Shows the wiring diagram of the setup for the SNTJ calibration. We send a $10$ Hz triangular voltage wave to bias the SNTJ through the DC port of a bias tee. The input signal goes through two $1.9$ MHz low pass filters one at room temperature and another at $40$ mK. The bias tee directs the noise power spectrum of the SNTJ to the output line. The noise spectrum is analyzed using a spectrum analyzer. The presence of the circulator is just to keep the output signal path identical for the SNTJ calibration and Gr-JPA noise measurement.
    \textbf{b,}~Shows the wiring diagram for the Gr-JPA measurements. The DC line for the gate passes through a $10$ Hz low pass RC filter followed by a $1.9$ MHz low pass filter at room temperature and a eccosorb filter at $40$ mK plate. The presence of the bias tee is just to keep the output signal path identical for the SNTJ calibration and Gr-JPA noise measurement.
    }}
\end{figure}

Full characterization of the Gr-JPA involves measurement of its gate tunability, nonlinear response, gain, and noise performance. We measure the noise performance from the signal-to-noise ratio improvement due to the Gr-JPA. The calibration of the system noise temperature (\Tns) and gain (\Gsys) of the amplification chain was done using the noise generated by a voltage-biased shot noise tunnel junction (SNTJ) in a separate cooldown. Next, during the measurements on the Gr-JPA, the SNTJ was replaced by the Gr-JPA to keep the amplification chain the same.\\
Supplementary Fig.~\ref{fig_1}a shows a schematic of the microwave setup for SNTJ calibration. The input line is made up of stainless steel coaxial cables and attenuated by $40$ dB using fixed attenuators at different temperatures stages. The output line consists of superconducting niobium-titanium coaxial cable from $40$ mK plate to the $4$ K plate and stainless steel cable for the rest. The output signal passes through a circulator (LNF) and two isolators (LNF) at the $40$ mK plate, one $40$ dB high-electron-mobility transistor (HEMT) amplifier at the $4$ K plate, and a $35$ dB room temperature amplifier. We use an arbitrary waveform generator (AWG, Agilent) to bias the SNTJ with a $10$ Hz triangular voltage signal and measure its shot noise spectrum using a spectrum analyzer (R\&S FSV signal analyzer). We use a bias tee (Inmet) to separate the path for DC and microwave signals. We choose two $1.9$ MHz low pass filters on the input line; one at room temperature and another one at $40$ mK plate to cut off any high-frequency noise.\\
Supplementary Fig.~\ref{fig_1}b shows a schematic of the microwave setup for measurements on the Gr-JPA. We have tried to keep the output signal path with respect to the device (SNTJ or Gr-JPA) as identical as possible with only an extra $6$ inch SMA coaxial cable and an SMA female-female adapter included with using the Gr-JPA, at the base temperature plate.
The input line is attenuated by $60$ dB using fixed attenuators at different temperatures stages. The output line comprises of two isolators (LNF) at the $40$ mK plate, one $40$ dB high-electron-mobility transistor (HEMT) amplifier at the $4$ K plate, and a $35$ dB room temperature amplifier. The signal from VNA (R\&S ZNB 20) combined with the pump (SignalCore RF generator) using a splitter (Mini-circuits) as combiner at room temperature is sent to the Gr-JPA. A magnetically shielded cryogenic circulator (LNF) is used to separate the incident and reflected signal from the Gr-JPA. The output signal is split using a splitter (Mini-circuits) at room temperature and is sent to the VNA and spectrum analyzer. The gate voltage ($V_{\mathrm{g}}$) is applied to the Gr-JPA using a DC voltage source (Keithley). The DC line for the gate passes through a $10$ Hz low pass RC filter followed by a $1.9$ MHz low pass filter at room temperature and an eccosorb filter at $40$ mK plate. The Gr-JPA is loaded inside a magnetic and radiation shielding cryogenic can inside our fridge.\\

\section{Details on noise measurements}
We measure the system noise temperature (\Tns) of our amplification chain with an SNTJ as a calibrated source of the noise. The SNTJ is a standard Al/AlO$_{\mathrm{x}}$/Al tunnel junction with $\sim60$ $\Omega$ normal state resistance ($300$ K value) which is embedded in a $50$ $\Omega$ CPW transmission line. A strong rare earth magnet is used to suppress superconductivity in Aluminium to realize a normal tunnel junction at $40$ mK with an estimated normal state resistance of $\sim67$ $\Omega$. The noise power coupled to a matched load by a voltage biased SNTJ at temperature $T$ and frequency $f$ is given by~\cite{spietz_primary_2003}
\begin{equation}\label{eqn:1}
    P_{\mathrm{n}}(f,V)=G_{\mathrm{sys}}(f)k_{\mathrm{B}}B\left[T_{\mathrm{n}}^{\mathrm{sys}}(f)+\frac{1}{2}\left(\frac{eV+hf}{2k_{\mathrm{B}}}\right)\coth{\left(\frac{eV+hf}{2k_{\mathrm{B}}T}\right)}+\frac{1}{2}\left(\frac{eV-hf}{2k_{\mathrm{B}}}\right)\coth{\left(\frac{eV-hf}{2k_{\mathrm{B}}T}\right)}\right]
\end{equation}
where \Gsys~is the gain of the amplification chain, $k_{\mathrm{B}}$ is the Boltzmann constant, $h$ is Planck's constant, $B$ is the bandwidth over which the noise spectra is measured, \Tns~is the system noise temperature of the amplification chain, and $V$ is the voltage applied across the tunnel junction. According to Eq.~\eqref{eqn:1}, measuring the noise power $P_{\mathrm{n}}(f,~V)$ one can determine the system temperature $T$, gain \Gsys,~and noise temperature \Tns~of the amplification chain. The high voltage limit of Eq.~\eqref{eqn:1} results in a linear relation between $P_{\mathrm{n}}$ and $V$ given by
\begin{equation}\label{eqn:2}
    P_{\mathrm{n}}(f,V)=G_{\mathrm{sys}}(f)k_{\mathrm{B}}B\left(T_{\mathrm{n}}^{\mathrm{sys}}(f)+\frac{e\abs{V}}{2k_{\mathrm{B}}}\right).
\end{equation}
Using Eq.~\eqref{eqn:2} we extract \Gsys ~and \Tns ~by a straight line fit of $P_{\mathrm{n}}(f,~V)$. We include a small correction ($\sim2\%$) to the noise power term in Eq.~\eqref{eqn:2} due to the SNTJ being slightly different than $50$ $\Omega$. Once the \Gsys ~and \Tns~of the amplification chain is known, the SNTJ is replaced by the Gr-JPA and we include a small correction to the \Gsys~and \Tns~due to an extra $6$ inch coaxial cable and SMA adapter with an estimated attenuation of $\sim0.1$ dB at $40$ mK.\\ 
When the Gr-JPA is biased for amplification, the system noise temperature (\Tnsjpa) of the amplification chain including the Gr-JPA is determined by measuring the signal-to-noise ratio improvement $\eta_{\mathrm{SNR}} = (S_{\mathrm{on}}/N_{\mathrm{on}})/(S_{\mathrm{off}}/N_{\mathrm{off}})$. 
Here, $S_{\mathrm{on/off}}$ represents the signal power when the Gr-JPA is on/off and $N_{\mathrm{on/off}}$ represents the noise floor power when the Gr-JPA is on/off. Finally, we extract the \Tnsjpa~using the following expression
\begin{equation}\label{eqn:3}
    T_{\mathrm{n}}^{\mathrm{sys,Gr-JPA}}(f)=\frac{T_{\mathrm{n}}^{\mathrm{sys}}(f)}{\eta_{\mathrm{SNR}}(f)}
\end{equation}

\subsection{SNTJ calibration}
We measure the SNTJ noise spectrum as a function of bias voltage. To do that we voltage bias the SNTJ using a $10$ Hz triangular wave; we use a spectrum analyzer in time-domain mode and capture the noise power in the half-cycle ($50$ millisecond) of the biasing triangular wave. Eventually, discarding the "time", we parametrically get the noise power as a function of bias voltage $P_{\mathrm{n}}(V)$. We measure the noise spectrum at different frequencies (in a band where the Gr-JPA works); hence, we get the frequency-dependent noise spectra $P_{\mathrm{n}}(V, f)$ of the SNTJ. Supplementary Fig.~\ref{fig_2}a shows a noise data at $f_{\mathrm{s}}=4.76$ GHz as a function of bias voltage at $40$ mK.

\subsection{Analysis of system noise temperature}
We fit the measured SNTJ noise power with a straight line using Eq.~\eqref{eqn:2}; shown in supplementary Fig.~\ref{fig_2}a. From the fitting parameters, we extract \Gsys ~and \Tns ~of the amplification chain including the correction due to SNTJ mismatch from $50$ $\Omega$ and extra attenuation of the extra $6$ inch coaxial cable and SMA adapter. The frequency-dependent \Gsys ~and \Tns~are plotted in supplementary Fig.~\ref{fig_2}b, c respectively. We use the numbers \Tns$(f)$ to extract the Gr-JPA noise performance. We use the \Gsys$(f)$ to calibrate the power level at the Gr-JPA, hence the $1$-dB compression point.

\begin{figure}[H]
    \centering
	\includegraphics[width=0.85\linewidth]{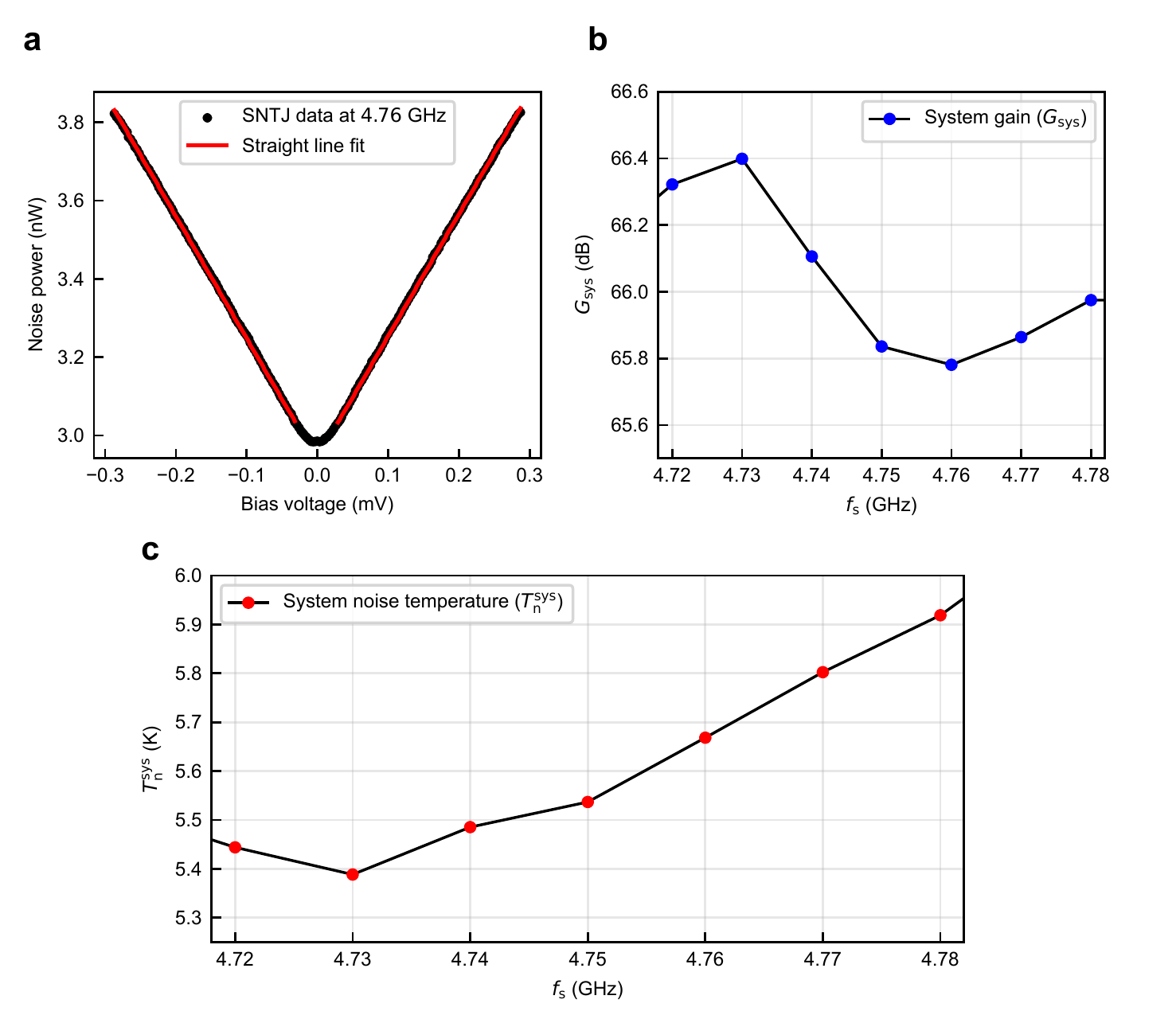} 
    \caption{ \label{fig_2} {\textbf{SNTJ calibration and system noise temperature \Tns extraction.} 
    \textbf{a,}~Shows the fitted noise spectra $P_{\mathrm{n}}(V)$ of the SNTJ at $4.76$ GHz, using Eq.~\eqref{eqn:2} in the high voltage limit at $40$ mK.
    \textbf{b,}~Shows the extracted system gain \Gsys~as a function of signal frequency.
    \textbf{c,}~Shows the extracted system noise temperature \Tns~as a function of signal frequency. 
    }}
\end{figure}

\subsection{Extraction of Gr-JPA noise temperature}
As discussed previously the Gr-JPA noise performance is determined by measuring the signal to noise ratio improvement $\eta_{\mathrm{SNR}} = (S_{\mathrm{on}}/N_{\mathrm{on}})/(S_{\mathrm{off}}/N_{\mathrm{off}})$~in presence of the Gr-JPA. We measure the signal power $S_{\mathrm{on/off}}$ when the Gr-JPA is on/off; shown in supplementary Fig.~\ref{fig_3}a. We next measure the noise floor power $N_{\mathrm{on/off}}$ when the Gr-JPA is on/off; shown in supplementary Fig.~\ref{fig_3}b. Next, we calculate the system noise temperature (\Tnsjpa) of the amplification chain including the Gr-JPA using Eq.~\eqref{eqn:3}; for \Tns$(f)$ we have used an average of the values extracted in the range $4.73$ to $4.77$ GHz which is the relevant amplification band. The error bars indicate $\pm$ one standard deviation of the \Tns~values. The measured \Tnsjpa~is plotted in supplementary Fig.~\ref{fig_3}c.

\begin{figure}[H]
    \centering
	\includegraphics[width=0.85\linewidth]{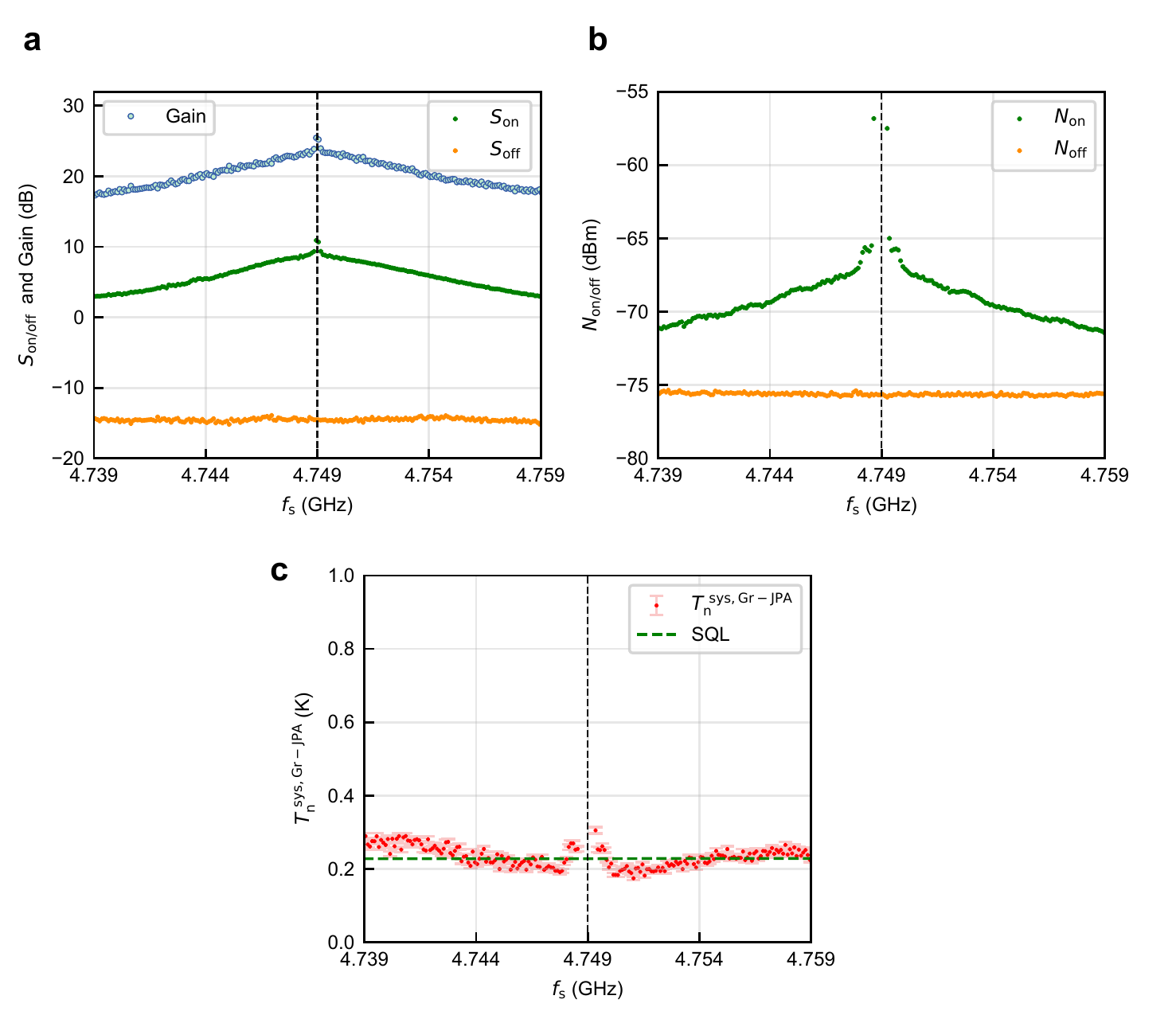}
    \caption{ \label{fig_3} {\textbf{Extraction of Gr-JPA noise temperature at $\bm{V_{\mathrm{g}}=4}$ V.} 
    \textbf{a,}~Shows the signal power $S_{\mathrm{on/off}}$ when the Gr-JPA is on/off. The difference of these two signals is the gain.
    \textbf{b,}~Shows the noise power $N_{\mathrm{on/off}}$ when the Gr-JPA is on/off.
    \textbf{c,}~Shows the system noise temperature of the amplification chain including Gr-JPA (\Tnsjpa). The green dashed line is the standard quantum limit (SQL).
    }}
\end{figure}

\section{Additional amplification data at different $V_{\mathrm{g}}$ values}
Here we show the Gr-JPA characterization data at different points $V_{\mathrm{g}}=3,~5,~7,~9,~11$ V. We explore the nonlinear phase response followed by the amplification of the Gr-JPA. We observe a doubly peaked gain profile at some gate biasing $V_{\mathrm{g}}=7,~9$ V (see supplementary Fig.~\ref{fig_6}). However, at other points, $V_{\mathrm{g}}=3,~5,~11$ V the gain profile is singly peaked and clean as the one at $V_{\mathrm{g}}=4$ V shown in the main manuscript. 

\subsection{Nonlinear phase diagram at different $V_{\mathrm{g}}$ points}
In supplementary Fig.~\ref{fig_4} we show the nonlinear phase diagram of the Gr-JPA at $V_{\mathrm{g}}=3,~5,~11$ V. The reflected phase ($\angle{S_{11}}$) is measured as a function of microwave signal frequency ($f_{\mathrm{s}}$) and power ($P_{\mathrm{s}}$).

\vspace{-0.58cm}
\begin{figure}[H]
    \centering
	\includegraphics[width=0.85\linewidth]{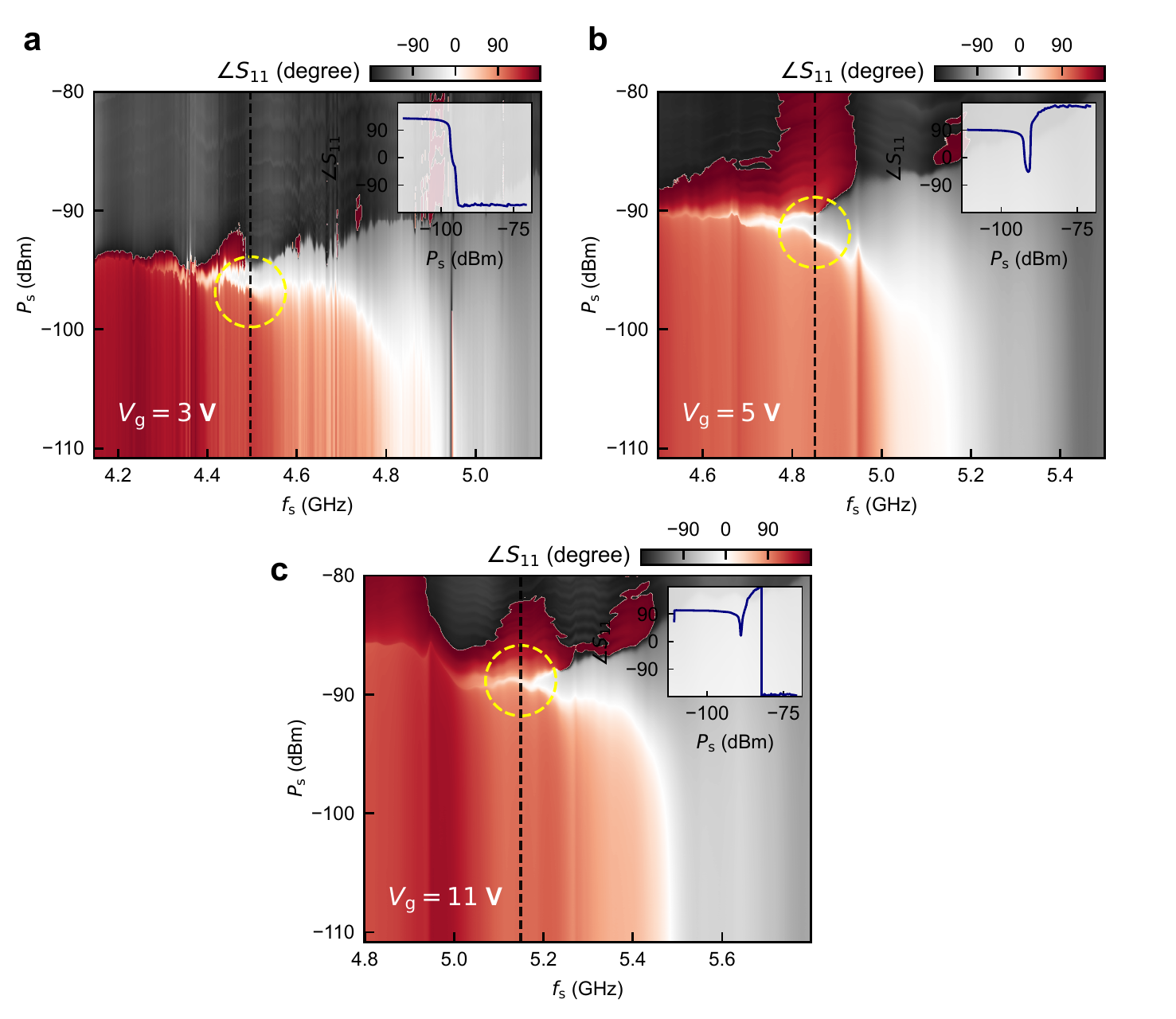} 
    \caption{ \label{fig_4} {\textbf{Nonlinear phase diagram at $\bm{V_{\mathrm{g}}=3,~5,~11}$ V.} 
    \textbf{a,}~Shows the reflected phase ($\angle{S_{11}}$) as a function of microwave signal frequency ($f_{\mathrm{s}}$) and power ($P_{\mathrm{s}}$) at $V_{\mathrm{g}}=3$ V. We mark a vertical dashed line on the frequency axis; upon introducing a pump tone ($f_{\mathrm{p}}$) at that frequency point we see amplification of probe signals. The yellow dashed circle indicates the bifurcation regime.
    \textbf{b,}~Shows the same phase measurement at $V_{\mathrm{g}}=5$ V.
    \textbf{c,}~Shows the same phase measurement at $V_{\mathrm{g}}=11$ V.
    }}
\end{figure}

\subsection{Gr-JPA amplification at different $V_{\mathrm{g}}$ points}
In supplementary Fig.~\ref{fig_5} we show the Gr-JPA amplification at $V_{\mathrm{g}}=3,~5,~11$ V respectively. At all these $V_{\mathrm{g}}$ points we have obtained at least $20$ dB amplification. In supplementary Fig.~\ref{fig_6}a, we show gain along with the noise temperature for the bias point $V_{\mathrm{g}}=3$ V. The noise performance varied from the SQL. We attribute this noise performance variation to be due to the variations in the environmental impedance. This is evident in supplementary Fig.~\ref{fig_4} where the effective resonance (white) doesn't vary as smoothly as in the simulated data in main manuscript Fig. 3b which uses a constant $50$ $\Omega$ impedance. Due to the same reason, we observe a double peak gain profile at $V_{\mathrm{g}}=7,~9$ V (supplementary Fig.~\ref{fig_6}b, c). Such variations are always present in real devices due to impedance mismatches~\cite{vijay_invited_2009}, and there is scope for improvement with better microwave design in future experiments.

\begin{figure}[H]
    \hspace*{-0.3cm}
    \centering
	\includegraphics[width=1.05\linewidth]{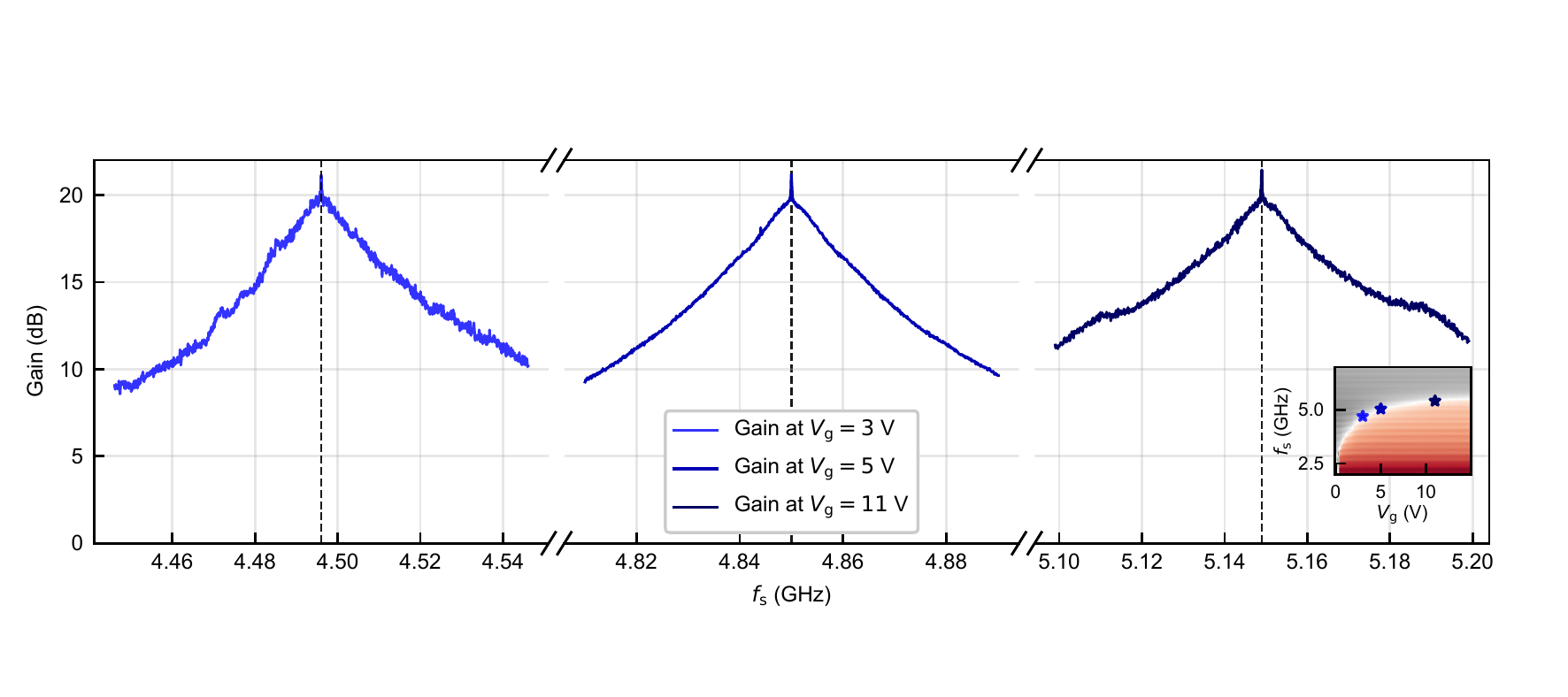} 
    \caption{ \label{fig_5} {\textbf{Amplification of the Gr-JPA at $\bm{V_{\mathrm{g}}=3,~5,~11}$ V.} 
    Shows the Gr-JPA gain at different $V_{\mathrm{g}}$ values. The $V_{\mathrm{g}}$ fixes the linear resonance ($f_{\mathrm{res}}^{\mathrm{lin}}$) of the Gr-JPA as indicated by the blue (light to dark) stars in the gate dispersion curve (inset bottom right). Introducing a pump tone near the bifurcation regime as marked in supplementary Fig.~\ref{fig_4} we see an amplification of at least $20$ dB at all these bias points. The position of the pump tone on the frequency axis is indicated by the black dashed vertical lines.
    }}
\end{figure}

\begin{figure}[H]
    \hspace*{-0.42cm}
    \centering
	\includegraphics[width=0.85\linewidth]{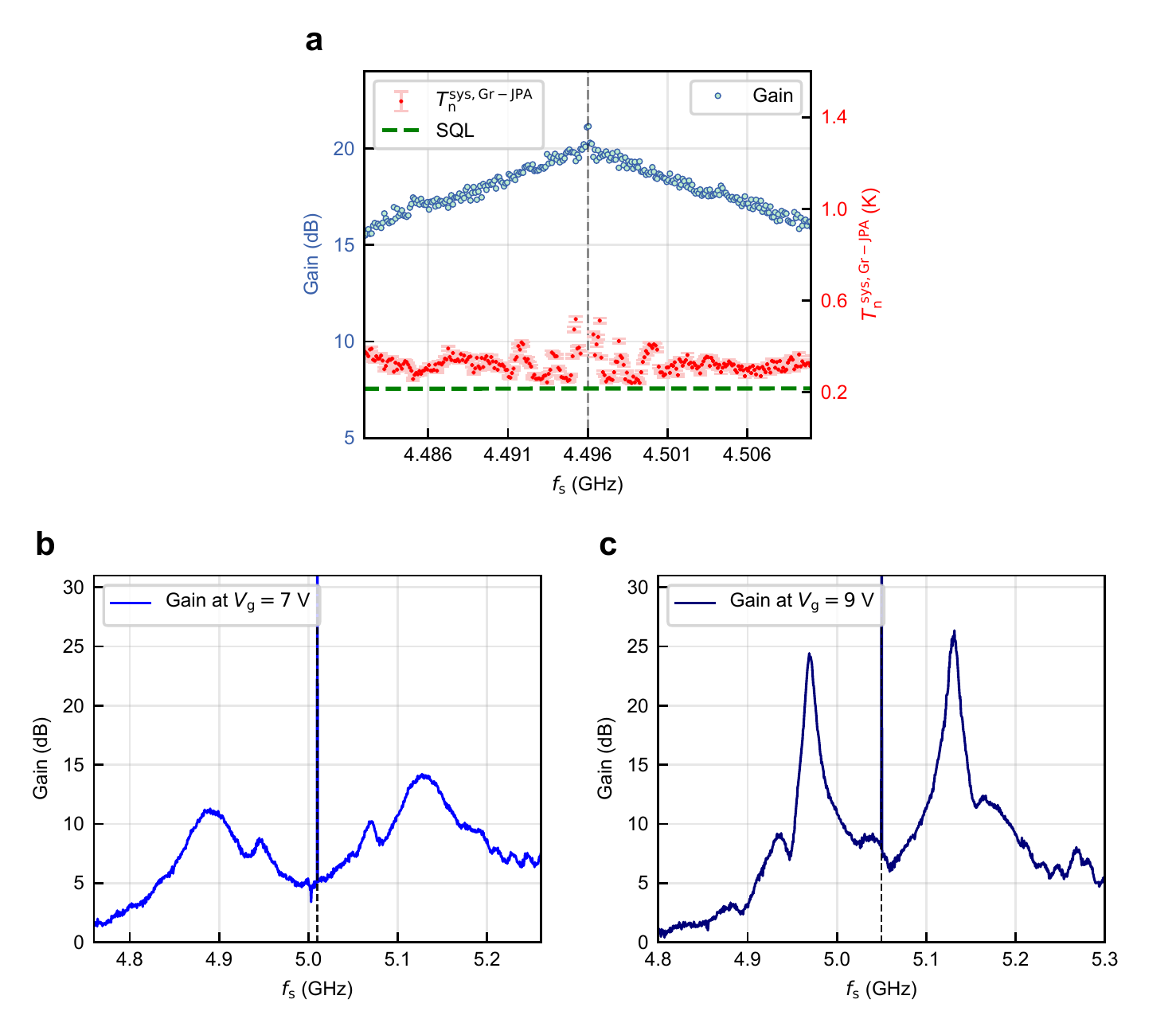}
    \caption{ \label{fig_6} {\textbf{Gr-JPA gain with noise temperature at $\bm{V_{\mathrm{g}}=3}$ V and doubly peaked gain profile at $\bm{V_{\mathrm{g}}=7,~9}$ V.} 
    \textbf{a,}~Shows gain along with the noise temperature (\Tnsjpa) at $V_{\mathrm{g}}=3$ V. The noise performance varies from the standard quantum limit (SQL) at some frequencies; primarily we attribute this due to environmental impedance mismatches. 
    \textbf{b,}~Shows a bias point at $V_{\mathrm{g}}=7$ V where the gain profile is doubly peaked around pump tone (black dashed vertical line) and the maximum gain is also low ($\sim15$ dB).
    \textbf{c,}~Shows a bias point at $V_{\mathrm{g}}=9$ V where the gain profile is doubly peaked around pump tone (black dashed vertical line) however, the maximum gain reaches $\sim25$ dB at some signal frequencies.
    }}
\end{figure}

\section{Additional data from another device}
We measured a few more devices. In supplementary Fig.~\ref{fig_7}a, b, c we show response of another device (Gr-JPA-2). We performed DC and microwave characterizations; which showed good tuning of critical current and hence linear resonance frequency.

\begin{figure}[H]
    \centering
	\includegraphics[width=0.85\linewidth]{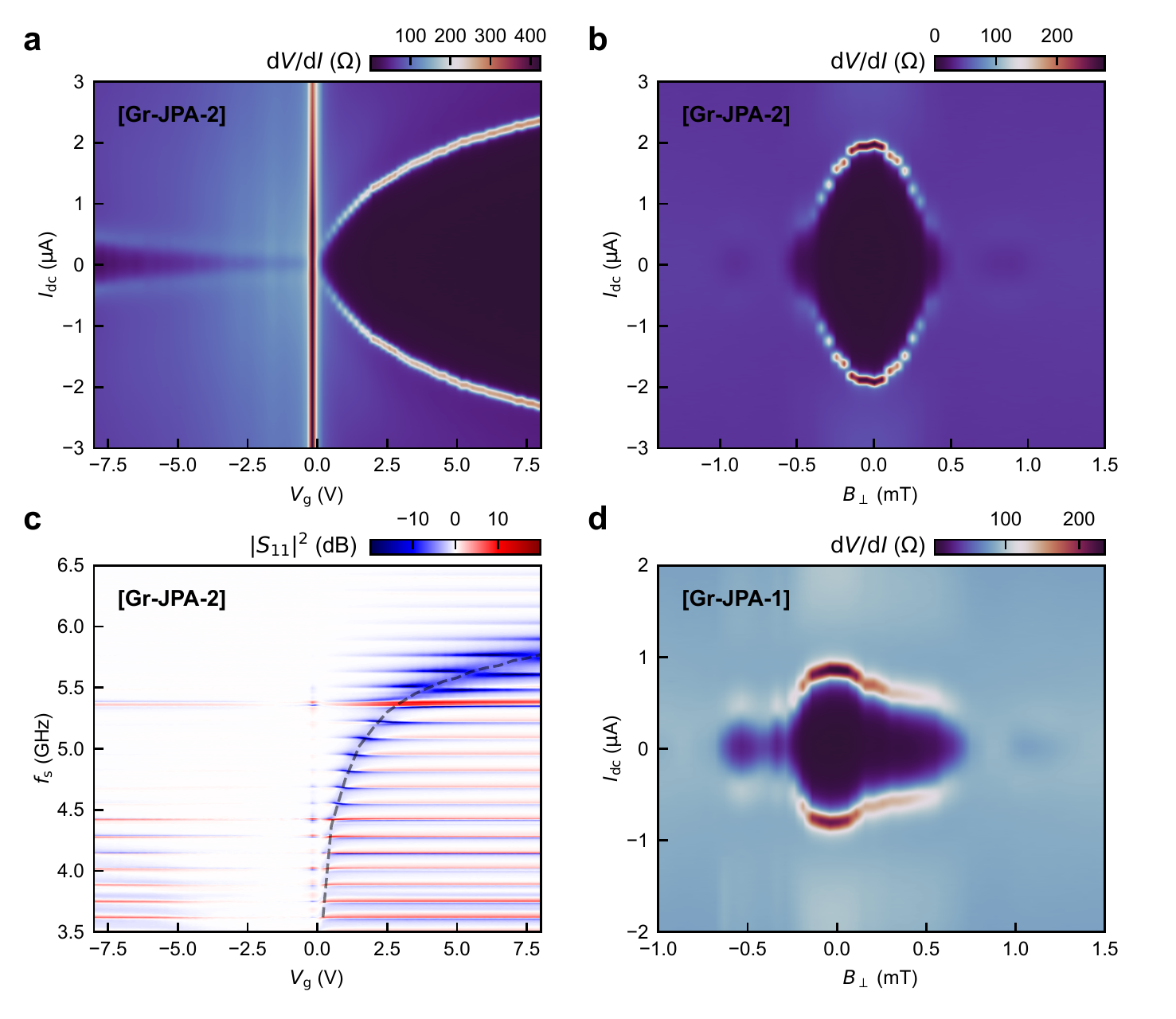} 
    \caption{ \label{fig_7} {\textbf{Additional data from another Gr-JPA device.} 
    \textbf{a,}~Shows differential resistance map ($\mathrm{d}V/\mathrm{d}I$) of Gr-JPA-2 device as a function of DC bias current ($I_{\mathrm{dc}}$) and applied gate voltage ($V_{\mathrm{g}}$). This device also show good tuning of critical current with gating.
    \textbf{b,}~Shows Fraunhofer modulation of the of Gr-JPA-2; differential resistance map ($\mathrm{d}V/\mathrm{d}I$) is plotted as a function of DC bias current ($I_{\mathrm{dc}}$) and applied perpendicular magnetic field. 
    \textbf{c,}~Shows the $\abs{S_{11}}^2$ (dB) of the Gr-JPA-2 as function of signal frequency ($f_{\mathrm{s}}$) and applied gate voltage ($V_{\mathrm{g}}$). Note: during this Gr-JPA-2 microwave measurements the circulator in the reflection measurement setup went bad and it was unable to provide desired isolation between through and isolated signal paths, and due to signal interference the ripples in the response appeared. However, the bending feature confirms that the linear resonance of the device is tuning as a function of gating.
    \textbf{d,}~Shows Fraunhofer modulation of the of Gr-JPA-1; differential resistance map ($\mathrm{d}V/\mathrm{d}I$) is plotted as a function of DC bias current ($I_{\mathrm{dc}}$) and applied perpendicular magnetic field. 
    }}
\end{figure}

\section{Details on microwave office simulations}
We extract the critical current dependence on gate voltage $I_{\mathrm{c}}$($V_{\mathrm{g}}$) from the $\mathrm{d}V/\mathrm{d}I$ map of main manuscript Fig. 2a. Following past work~\cite{titov_josephson_2006} on modeling the supercurrent in gr-JJs, we model the CPR using a simplified relation, 
\begin{equation}\label{eqn:4}
    I_{\mathrm{s}}(\phi) = \frac{\pi\Delta_0}{2eR_{\mathrm{n}}} \frac{\sin{\phi}}{\sqrt{1-\tau\sin^2{(\phi/2)}}}
\end{equation}
Where $I_{\mathrm{s}}$ is the supercurrent, $\phi$ is the phase difference across the junction, $\Delta_0$ is the induced superconducting energy gap in graphene, $R_{\mathrm{n}}$ is the normal state resistance of the junction and $\tau$ is the averaged transparency factor for $N$ conducting channels in graphene. Eq.~\eqref{eqn:4} has three parameters $\Delta_0$, $R_{\mathrm{n}}$, $\tau$. The maximum value of $I_{\mathrm{s}}(\phi)$ with respect to $\phi$ is the critical current ($I_{\mathrm{c}}$) of the junction. We know $I_{\mathrm{c}}$, $R_{\mathrm{n}}$ from experimental data. We estimate the $\Delta_0$ from $I_{\mathrm{c}}R_{\mathrm{n}}$ product. We use the limiting values of $\Delta_0/e=I_{\mathrm{c}}R_{\mathrm{n}}/2.08$ at CNP ($V_{\mathrm{g}}\sim0$ V) and $\Delta_0/e=I_{\mathrm{c}}R_{\mathrm{n}}/2.44$ at far away from CNP ($V_{\mathrm{g}}\gg 0$ V) for short ballistic junctions~\cite{titov_josephson_2006}. In between this two limits we assume a linear approximation of the prefactor ($p_f$) and assume the form $\Delta_0/e=I_{\mathrm{c}}(V_{\mathrm{g}})R_{\mathrm{n}}(V_{\mathrm{g}})/p_{f}$.\\
We numerically calculate $\tau$ by equating the maxima of Eq.~\eqref{eqn:4} to $I_{\mathrm{c}}$ for each $V_{\mathrm{g}}$ values using a search and match method. As a result we extract $\tau$($V_{\mathrm{g}}$) as shown in supplementary Fig.~\ref{fig_8}a. Once we know $\tau$($V_{\mathrm{g}}$) the CPR at each $V_{\mathrm{g}}$ is known. Next we take numerical derivative of the CPR with respect to $\phi$ to calculate the $L_{\mathrm{J}}$($\phi$) as,
\begin{equation}
    L_{\mathrm{J}}(\phi) = \frac{\hslash}{2e} \left(\frac{\partial I_{\mathrm{s}}}{\partial\phi}\right)^{-1}.
\end{equation}
Numerically calculated $L_{\mathrm{J}}$($\phi$) is plotted in supplementary Fig.~\ref{fig_8}b at $V_{\mathrm{g}}=4$ V. Next, we parametrically extract $L_{\mathrm{J}}$($I$) from supplementary Fig.~\ref{fig_8}b. Eventually, we fit the extracted $L_{\mathrm{J}}$($I$) with a polynomial and get the nonlinear inductance functional form: $L_{\mathrm{J}}(I)=L_0+L_1I+L_2I^2+L_3I^3+..$ $L_{\mathrm{J}}$ as a function of ($I_{\mathrm{s}}/I_{\mathrm{c}}$) is shown in supplementary Fig.~\ref{fig_8}c.\\
Next, we model a nonlinear inductor in the AWR microwave office using the coefficients $L_{i}^{~,}$s and simulate the nonlinear phase diagram of an LC resonator as a function of microwave signal frequency and power. The inductance value used in microwave simulation is the sum of $L_{\mathrm{J}}$ and $L_{\mathrm{stray}}$~as $L_{\mathrm{J}}(I)=(L_0+L_{\mathrm{stray}})+L_1I+L_2I^2+L_3I^3+..$~We use the value of experimentally determined linear resonance frequency ($f_{\mathrm{res}}^{\mathrm{lin}}$) and the linear Josephson inductance ($L_0$) calculated at the operating point ($V_{\mathrm{g}}=4$ V) to fit our circuit model with an estimated value of $L_{\mathrm{stray}}=50$ pH  and a total shunt capacitance of $1.219$ pF. We use the APLAC Harmonic Balance (HB) technique for the simulation. One simulated phase diagram is shown in the main manuscript Fig. 3b.

\begin{figure}[H]
    \centering
	\includegraphics[width=0.85\linewidth]{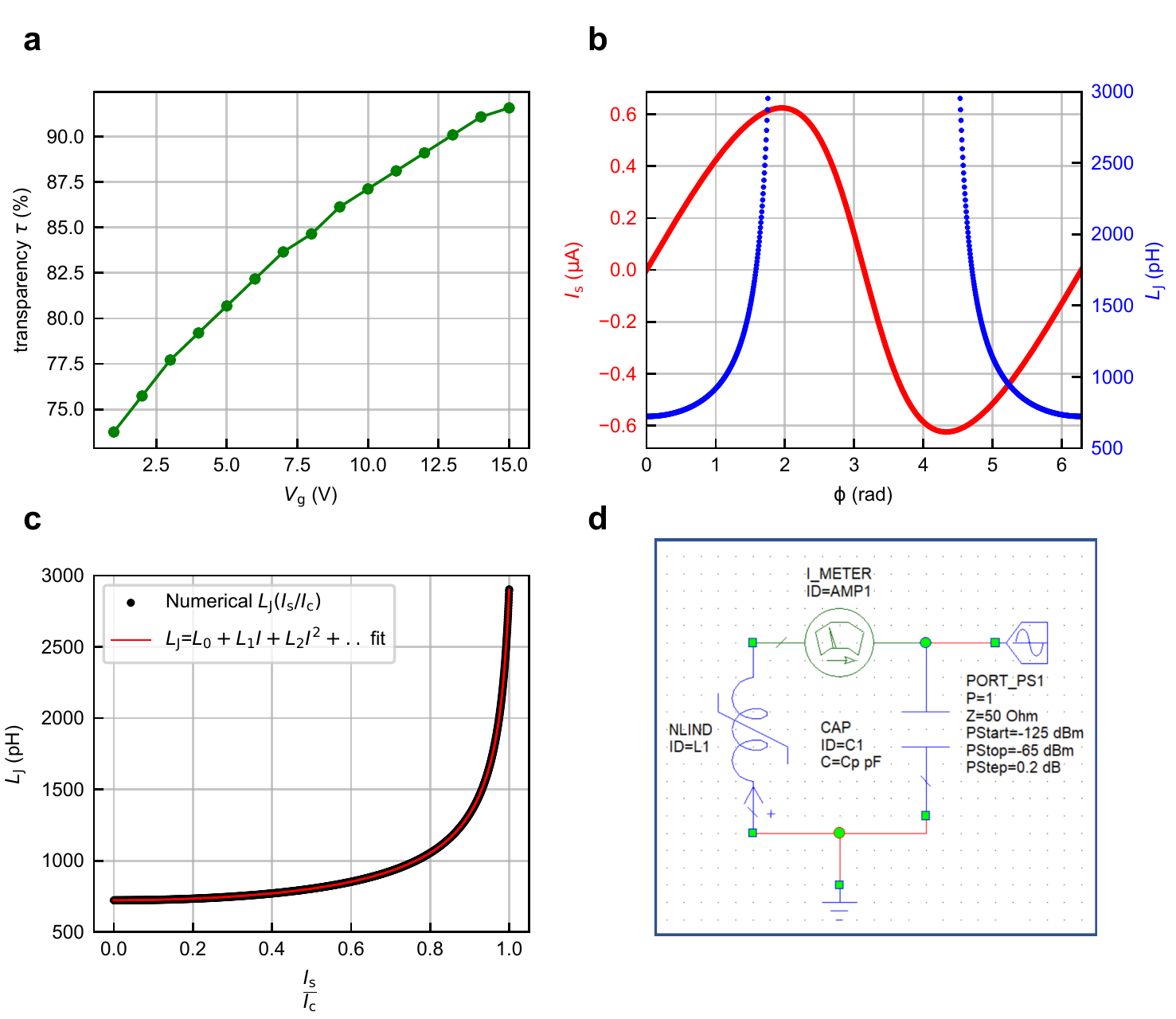} 
    \caption{ \label{fig_8} {\textbf{Numerical simulations.} 
    \textbf{a,}~Shows the JJ transparency ($\tau$) as a function of $V_{\mathrm{g}}$ that we extract from the main manuscript Fig.2a data. The $\tau$ extraction is based on the CPR (Eq.~\eqref{eqn:4}). 
    \textbf{b,}~Shows a numerically extracted CPR and junction inductance ($L_{\mathrm{J}}$) as a function of phase ($\phi$) at $V_{\mathrm{g}}=4$ V.
    \textbf{c,}~Shows a plot of junction inductance as a function of normalized current ($I_{\mathrm{s}}/I_{\mathrm{c}}$) at $V_{\mathrm{g}}=4$ V. We extract this from the Fig.\ref{fig_8}b; next we fit a polynomial and extract the nonlinear inductor coefficients of $L_{\mathrm{J}}(I)=L_0+L_1I+L_2I^2+L_3I^3+..$. 
    \textbf{d,}~Shows a circuit schematic in the microwave office that we use for the nonlinear phase diagram simulation. The "NLIND" element is the nonlinear inductor $L_{\mathrm{J}}(I)=(L_0+L_{\mathrm{stray}})+L_1I+L_2I^2+L_3I^3+..$
    }}
\end{figure}

\end{document}